# Electrodynamics with a Future Conformal Horizon

## Michael Ibison

*Institute for Advanced Studies at Austin*
*11855 Research Boulevard, Austin TX 78759, USA*

**Abstract.** We investigate the impact of singularities occurring at future times in solutions of the Friedmann equations expressed in conformal coordinates. We focus on the consequences of extending the time coordinate through the singularity for the physics of matter and radiation occupying just one side. Mostly this involves investigation of the relationship between the metric with line element $ds^2 = a^2(t)(dt^2 - d\mathbf{x}^2)$ and time reversal symmetry within electrodynamics. It turns out compatibility between these two is possible only if there is a singular physical event at the time of the singularity or if the topology is not trivial. In both cases the singularity takes on the appearance of a time-like mirror. We are able to demonstrate a relationship between the broken time symmetry in electrodynamics characterized by retarded radiation and radiation reaction and the absolute conformal time relative to the time of the singularity, i.e. between the Electromagnetic and Cosmological arrows of time. It is determined that the Wheeler-Feynman reasoning but with the future absorber replaced by the Cosmological mirror leads to a conflict with observation unless matter is electromagnetically strongly bound to the environment.

## INTRODUCTION

If expressed in conformal time the scale factor that solves the first Friedmann is singular in the finite future. Generally this is not taken too seriously because the singularity occurs in the infinite future according to the time of the Friedmann-Robertson-Walker (FRW) coordinate system, generally taken to be the time recorded by laboratory clocks. Further, from the perspective of GR the singularity is a consequence of the particular choice of coordinate system and can be traversed with a suitable redefinition. To ignore the singularity seems questionable from the standpoint the electromagnetic fields, however. The Maxwell theory is conformally invariant, and can be cast so the fields and associated potentials – appropriately defined – are completely insensitive to the singularity [1]. In that case, one might presume they pass through the singularity as if it was not there. Keeping in mind that due to conformal invariance light is neither attenuated nor red-shifted by the time it arrives as the singularity, it seems especially urgent to determine its fate. Where does it go? Does this not mean that spacetime after the time of the singularity should be granted the same existential status as the spacetime before it?

If time is allowed to continue through the singularity, it follows there must exist a post-singularity universe containing real matter and radiation (at least the matter and radiation having crossed over from this side). Taking into account the anti-symmetry of the scale factor about the singular time, that universe must in some sense be a mirror image of our own. In the following we show how this leads to the promotion of the singularity to a boundary condition. Considered in turn are the consequences for classical matter, the Dirac wavefunction, and electromagnetic fields, respectively. The boundary causes the intrinsic time-symmetry of the Maxwell theory to be broken locally. We investigate the relationship between this and the electromagnetic arrow of time manifesting as exclusively retarded radiation and radiation reaction.

# THE CONFORMAL BOUNDARY

Cosmological expansion is accommodated by a flat-space FRW metric whose line element can be written [1]

$$ds^2 = d\tau^2 - a^2(\tau)d\mathbf{x}^2 \tag{1}$$

where $\tau$ is the proper time of a co-moving (fundamental) observer and $a$ is the FRW scale factor. In these coordinates the coordinate density of a dilute matter fluid is constant whilst the coordinate density of EM energy falls as $\rho_{EM}(0)/a(\tau)$.[2] The latter behavior is responsible for Cosmological red-shift with the wavelength of light increasing linearly with expansion. With the definition of a new time [3]

$$dt = d\tau/a(\tau) \tag{2}$$

the line element (1) becomes

$$ds^2 = a^2(t)dx^2 \tag{3}$$

where $dx^2 = dt^2 - d\mathbf{x}^2$ (implying the implicit definition of the *Lorentz* vector $\{x^\mu\} = (t,\mathbf{x})$ with a *Minkowski* metric). In these *conformal* coordinates the invariant interval differs from that of Minkowski space-time only by a factor. A consequence is that the electromagnetic influence of a point source is no different in these coordinates than in a space that is not expanding at all.[4] Depending on the choice of gauge (see below) not only the light cone but the Maxwell's equations in general can be rendered insensitive to the scale factor, making these coordinates useful for electrodynamics calculations on the large scale. Important for the subsequent discussion is the property of conformal coordinates that they may perhaps index parts of the manifold that are inaccessible to the FRW coordinates. In particular, and as easily deduced from the definition (2), if the scale factor increases faster than linear in FRW time then $\tau = \infty$ occurs at a finite $t$ time.

The evolution of the scale factor is decided by GR. Post recombination this boils down to the Friedmann equation plus equations of state for the various contributions.[5] In the FRW system the former is [2]

$$\frac{a}{H^2}\left(\frac{da}{d\tau}\right)^2 = \frac{\Omega_{EM}}{a} + \Omega_m + \Omega_\Lambda a^3 \tag{4}$$

where the overall factor has been chosen so that each term is a *coordinate* energy density. Each of the $\Omega$ is an energy density normalized so that their sum is unity, and $\Lambda$ denotes the vacuum contribution. Present estimates are [3]

$$H^{-1} = 9.78/.72 = 13.6 \text{ Gyr}$$
$$\Omega_\Lambda = 0.74, \quad \Omega_m = 0.256, \quad \Omega_{EM} = 4.76 \times 10^{-5} \tag{5}$$

It is clear from inspection that solutions to (4) will expand indefinitely and exponentially once the vacuum term starts to dominate. That it ever does so requires that some minimum condition is satisfied involving the initial condition and the relative magnitude of the $\Omega$, which condition turns out is easily met in our universe [2]. In the late phase then

$$a(\tau) \approx e^{H\tau} \tag{6}$$

where, following convention, the scale factor is set to unity at the present time $\tau = 0$.

The Friedmann equation in conformal coordinates is simply (4) with (2)

$$\frac{1}{H^2}\left(\frac{da}{dt}\right)^2 = \Omega_{EM} + \Omega_m a + \Omega_\Lambda a^4 \tag{7}$$

where, as in (4), each term is a coordinate energy density. The vacuum-dominated asymptotic behavior inferred from this is that of a simple pole at the boundary

$$a(t) = (1 - Ht)^{-1} \tag{8}$$

---

[1] Here and throughout $c = 1$.

[2] The proper densities fall respectively as $\rho_m(0)/a^3(\tau)$ and $\rho_{EM}(0)/a^4(\tau)$.

[3] The conformal time is that recorded e.g. by a light clock where each tick is the bounce of a light pulse between two parallel mirrors co-moving with the Hubble expansion. By contrast the FRW time is that recorded by an appropriately defined laboratory clock 'not subject to the large scale expansion'.

[4] The light cone, a surface of co-dimension 1 in 3+1D and hence a hyper-surface of dimension 3, is invariant under expansion.

[5] Either the equations of state or the second Friedmann equation involving the pressure.

where the time may be presumed zero at, and start with, the big bang, but is open at the other end: $t \in \mathbb{R}_+$.[6] Henceforth we will refer to the 3-surface $\forall \mathbf{x}, t = 1/H$ as the 'conformal boundary', and the regions either side as pre and post boundary and, more informally, as lower and upper, half-spaces. We live in the lower half-space.[7] Using (2) with (8) one has

$$\tau = \int_0^t dt' (1 - Ht')^{-1} = -H^{-1} \log|1 - Ht|; \quad t \neq 1/H \tag{9}$$

which demonstrates that the domain $t \in \mathbb{R}_+$ is mapped twice to that of $\tau$ everywhere except at $t = 1/H$.[8]

From the above it may be concluded that a solution of the Friedmann equation in FRW time followed by a coordinate transformation to conformal time is not generally the same as a solution of the Friedmann equation in conformal time. All cosmologies compatible with observation evolve asymptotically in conformal time as (8). Note however that though the scale factor is singular at the conformal boundary it does not follow - and has yet to be determined - that anything strange or dramatic happens to matter (including EM fields) there. That will be the focus of the sections following.

The above is valid only asymptotically. A more accurate calculation of the conformal time to the future boundary requires integration of (7). With the numbers (5) numerical integration gives

$$t(a = \infty) - t(a = 1) = \frac{1}{H} \int_1^\infty \frac{da}{\sqrt{\Omega_{EM} + \Omega_m a + \Omega_\Lambda a^4}} = 1.12/H = 15.23 \text{ Gyr} \tag{10}$$

At other than later epochs Eq. (8) is inappropriate (unqualified, (8) would predict zero scale only in the infinite past). At earlier times the scale factor must come from solution of the full Friedmann equation, which is a good approximation only for times later than the end of the recombination era, which occurred when the temperature fell to about 3000 K, and possessed therefore a scale factor around 0.1% of its size now.[9] Approximately the conformal time elapsed since then is

$$t(a = 1) - t(a = 0) = \frac{1}{H} \int_0^1 \frac{da}{\sqrt{\Omega_{EM} + \Omega_m a + \Omega_\Lambda a^4}} = 3.47/H = 47.19 \text{ Gyr} \tag{11}$$

which puts the present at around 75.6% of the total lifetime. The earlier development is beyond the scope of this article. Calculations taking into account evolution during the pre-recombination epochs, including inflation and baryonic, give comparatively significant contributions to the conformal age, changing therefore the fractional age of the present. Importantly, currently favored models of early evolution agree that the conformal age is *finite* and remains of order of $1/H$ [4].

## CLASSICAL MATTER

The classical inertial action is [5]

$$I = -m \int \sqrt{dx^\mu dx^\nu g_{\mu\nu}(x)} \tag{12}$$

where the integration is along the path of the particle, $\mathbf{x}(t)$ say. The metric corresponding to (3) is $g_{ab}(x) = a^2(t) \eta_{ab}$, so

$$I = -m \int dt |a(t)| \sqrt{1 - \mathbf{v}^2(t)}; \quad \mathbf{v}(t) := d\mathbf{x}(t)/dt \tag{13}$$

The geodesics of classical particles are therefore

---

[6] Eqs. (9) and (11) are such that the FRW and conformal clocks are presently synchronized $t = 0 \Leftrightarrow \tau = 0$ and tick at the same rate and with the same sense: $d\tau/dt\big|_{t=0} = 1$

[7] Subsequently that we live in either one or both will be less clear.

[8] If (2) is used with (6) then one obtains (8) through a map $\tau \to t$ that is valid only as $\mathbb{R} \to (-\infty, H^{-1}]$.

[9] Here we have reverted to the FRW picture in which EM radiation loses energy whilst co-moving matter is constant during expansion in accord with the terms in (4). In this picture the present CMB at 2.7K must have been at temperature 2700 K when $a$ was 1/1000. By contrast the conformal picture is that the CMB has not changed but that matter has since become more energetic, behaving as if the Compton frequency has risen in proportion to the scale factor. Accordingly when $a$ was 1/1000 the energy of recombination would have been 1000 times smaller than its present value of 0.25 eV, and associated therefore, with a temperature 1000 times less than 2700 K.

$$\frac{d}{dt}\frac{|a(t)|\mathbf{v}(t)}{\sqrt{1-\mathbf{v}^2(t)}} = \mathbf{0} \Rightarrow \mathbf{v}(t) = \frac{\mathbf{k}\sqrt{1-\mathbf{v}^2(t)}}{|a(t)|} \quad (14)$$

for some constant **k** - which can be expressed in terms of an initial velocity

$$\mathbf{k} = \mathbf{v}_0/\sqrt{1-\mathbf{v}_0^2} \quad (15)$$

where $\mathbf{v}_0 \equiv \mathbf{v}(0)$. With this (14) is

$$\mathbf{v}(t) = \mathbf{v}_0 \Big/ \sqrt{a^2(t) + \mathbf{v}_0^2(1-a^2(t))} \quad (16)$$

and with (8) in particular this is

$$\mathbf{v}(t) = \mathbf{v}_0 |1-Ht| \Big/ \sqrt{1 + \mathbf{v}_0^2((1-Ht)^2 - 1)} \quad (17)$$

Integrating again, the geodesics are found to be

$$\mathbf{x}(t) = \mathbf{x}_0 + \frac{\mathbf{v}_0}{H\mathbf{v}_0^2}\begin{cases} 1 - \sqrt{1 + \mathbf{v}_0^2((1-Ht)^2 - 1)}; & t \leq 1/H \\ 1 + \sqrt{1 + \mathbf{v}_0^2((1-Ht)^2 - 1)} - 2\sqrt{1-\mathbf{v}_0^2}; & t \geq 1/H \end{cases} \quad (18)$$

The Hubble flow is just the case that $\mathbf{v}_0 = \mathbf{0}$, in which case $\mathbf{x}(t) = \mathbf{x}_0$. Nothing peculiar happens at the conformal boundary; the matter just continues on as if the boundary were not there. The conformal diagram is given in Fig. 1. The Big Bang has been given a nominal conformal age of $-1.5/H$ though its actual value must be less than this. Here and henceforth in all figures all distances will be in units of $1/H$.

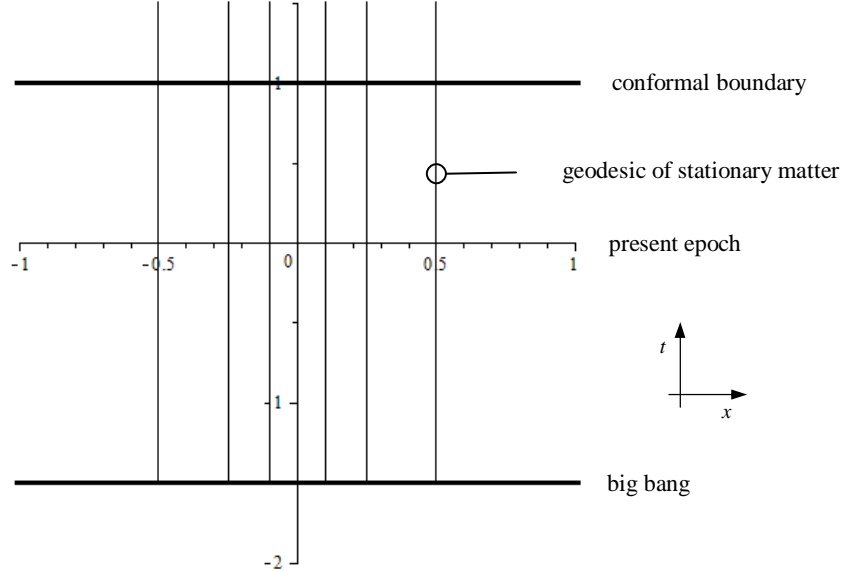

**Figure 1.** Typical geodesics of stationary matter in the conformal system. Distances are in units of $1/H$.

Figure 2 demonstrates the behavior of geodesics obeying (18) having an initial non-zero speed. Effectively the Hubble frame exerts a drag on inertial matter. Speeds are reduced to zero at the conformal boundary by gradual deceleration from their initial value at $t = 0$. Memory of the state immediately prior to the boundary is retained through the higher derivatives. Speeds increase after passing through, eventually reaching their initial value at a future time that is twice the initial distance from the boundary, i.e. symmetrically opposite the initial time. Accordingly, the present speed of a completely uninterrupted geodesic is the continuously evolving outcome of previous deceleration from historically higher speeds. An exception is for light-speed 'matter'; with reference to (16) we see that

$$\lim_{a \to \infty} |\mathbf{v}(t)| = \begin{cases} 0 & \text{if } |\mathbf{v}_0| < 1 \\ 1 & \text{otherwise} \end{cases} \quad (19)$$

The special case is not as singular as it might at first seem. The time spent in any fixed interval of speeds around |**v**| = 0 is progressively smaller for the initially faster moving particles. Hence one may alternatively regard the light-speed case as conforming to the general behavior, though spending zero time at zero speed at the boundary. At the other end one has

$$\lim_{a \to 0} |\mathbf{v}(t)| = 1 \tag{20}$$

so all geodesics were once asymptotically light-like at the time of the big bang, regardless of the subsequent model-dependent development of the scale factor.

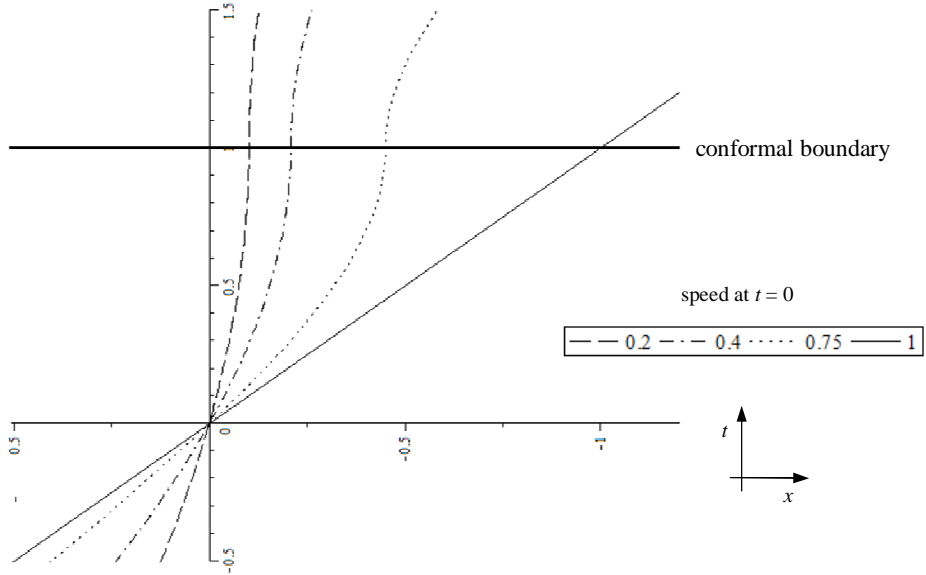

**Figure 2.** Hubble drag on geodesics passing through the origin for a range of initial speeds.

If the conformal coordinate $t$ is taken as an index into a real spacetime as $t \in \mathbb{R}_+$ then the Friedmann equation must be obeyed post horizon. The Friedmann equation dictates that the scale factor evolve post boundary in a way that is symmetric (or anti-symmetric) to the pre-boundary evolution. If so, the post boundary development of matter and radiation for example must be such as to explain an initially exponential *contraction* of the scale factor, passing through an era of re-ionization about 13 billion years later, (measured in conformal time) culminating in a big crunch. Though the scale factor is a gross approximation to the fine details, the story must remain the same at finer levels lest there be some inconsistency. In short, if the scale factor is anti-symmetric about the boundary, the configurations of matter and radiation must likewise be in some sense symmetric also.

Figure 2 as it stands cannot be correct therefore. If regarded as an ensemble of particles leaving the origin at $t = 0$, the pre and post boundary developments of the whole are not related by any simple symmetry operation. If we view matter as one of the 'causes' behind the scale factor development, as time progresses the scale factor so-driven would not retrace its steps as prescribed by the Friedmann equation. Consistency with the latter demands the geodesics be somehow mirrored in the conformal boundary. There are two distinct possibilities, as follows:

One possibility is that a mirror universe restores the symmetry so the boundary appears symmetrically both as a sink and a source. This is illustrated in Figure 3. It is equivalent to regarding the particles as perfectly reflected *in time* at the boundary, though from the familiar viewpoint of development in monotonic time it will appear as if all particles are annihilated by perfectly synchronized partners that are members of the *same* half-space. It is important to keep in mind that the mirror is time-like, and though Figure 3 appears to show reflections at the boundary, each particle and its nemesis collide head-on *in space*.

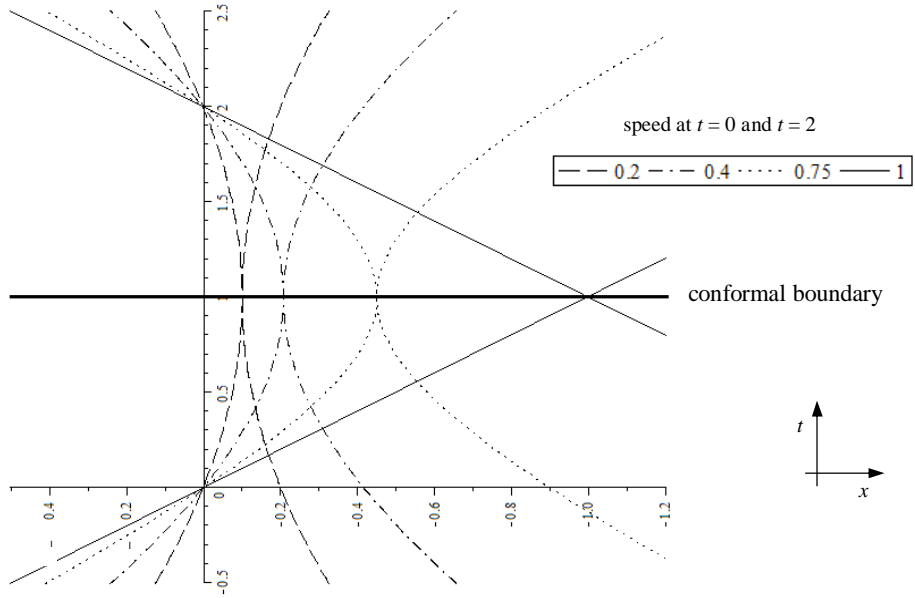

**Figure 3.** Geodesics of Figure 2 with reflections.

The other possibility is that the neutral classical particles can be made to conform to Fig. 4 through the action

$$I = -m \int dt\, a(t) \sqrt{1 - \mathbf{v}^2(t)} \tag{21}$$

which now replaces (13). Then the velocity evolves as

$$\mathbf{v}(t) = a(t)\mathbf{v}_0 \Big/ \sqrt{1 + \mathbf{v}_0^2 \left(1/a^2(t) - 1\right)} \tag{22}$$

and in the particular case of (8) the geodesics - with distances are normalized to the Hubble constant - are

$$\mathbf{x}(t) = \mathbf{x}_0 + \left[1 - \sqrt{1 + \mathbf{v}_0^2\left((t-1)^2 - 1\right)}\right] \mathbf{v}_0 / \mathbf{v}_0^2; \quad t \in \mathbb{R}_+ \tag{23}$$

The geodesics in Fig. 4 have the same initial configuration as in Fig. 2, but are now annihilated at the boundary.

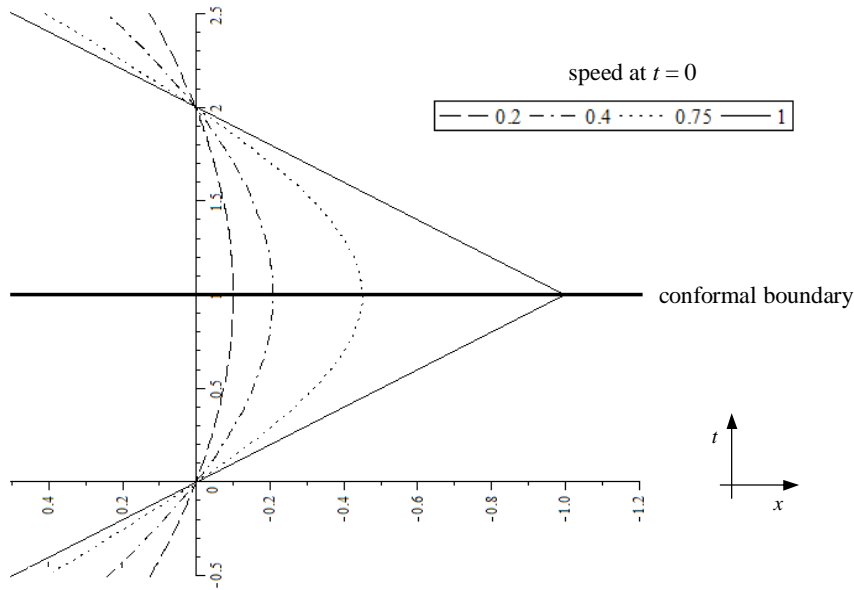

**Figure 4.** Geodesics of Fig. 2 annihilated by their image.

## Mirror or Topology?

With either possibility the distribution of classical neutral matter is not contradicted if the domain of the Friedmann equation if extended beyond the boundary. The conformal boundary appears as a time-like mirror in which the pre-boundary world-lines are time-reversed. If the *image* in the mirror is compatible with real (pre-boundary) physics then the two might even be regarded as co-existing and the boundary an artifact of projection. In the vicinity of the boundary external Lorentz forces can be ignored due to the effect of the scale factor, which makes the mass look infinite. We will see below that EM fields are affected differently by (or near) the boundary, but in a manner that leaves it relatively powerless to affect the trajectory of the increasingly massive particle from that illustrated here. Consequently, asymptotically all particles conform to the geodesics (23).

A real mirror requires physical matter to effect the reflection. The image in the mirror need not conform to physical law governing the 'real' side of the mirror. For example the notion of an 'image charge' employed to explain the behavior of an electron near a mirror was not taken, historically, as immediate proof of the necessary independent existence of a positive charge with the same mass. Likewise there need be no real universe on the other side of the time-like mirror in Fig. 3. The image need not be authentic therefore. In any case there must be some - as yet unidentified - physical process localized to the time of the Conformal boundary and which is responsible for the behavior illustrated there. In short: Figure 3 gives a possible arrangement whereby the Friedmann equation is not contradicted, but does not provide a physical explanation for the behavior at the boundary.

Another possibility, depicted in Figure 4, is that the symmetry is upheld without recourse to external world lines in the lower half-space. In this scenario the fate of a particle at the boundary is knowable locally. There is a continuous evolution through the boundary with no scattering or 4-particle vertex characteristic of the time-like reflections in Figure 3. This boundary has the character of being a consequence of a closed topology, where the coordinatization is hiding an intrinsic periodicity.[10] This possibility will be preferred in this document because it does not challenge existing observation with the requirement of orchestration of pairs destined for annihilation or coherent fields destined for cancellation. It does however come with tighter constraints. These include that the post-boundary universe is physically authentic, permitted by the pre-boundary laws of physics and their known symmetries. The relationship between this and the topology of the whole is an interesting topic that is explored only briefly here. The mirror versus topology issue will have to be re-visited when considering radiation.

The strong distinction between the two possibilities illustrated in Fig. 3 and Fig. 4 applies to classical matter, but is weakened when consideration of the boundary behavior is extended to *fields*. It seems to remain meaningful at the level of Dirac matter (below), but may be lost entirely at the level of quantized fields. It will be noticed that initially it seems meaningful to apply the distinction to classical radiation, though subsequently it becomes less well defined.

We will avoid a proper discussion of entropy in this document; all the possibilities considered herein are likely to do great violence to the (presumed) entropic arrow of time.[11]

## DIRAC WAVEFUNCTION

Development of the wavefunction in curved spacetime requires determination of a non-unique vierbein which generates an ambiguity in the fate of the geodesics of quantum matter. This ambiguity turns out to be related to the two possibilities discussed above. In the case of spacetime (3) one seeks a matrix $V$ such that [1]

$$g_{\mu\nu}(x) = V^{\alpha}{}_{\mu}(x) V^{\beta}{}_{\nu}(x) \eta_{\alpha\beta} = a^2(t) \eta_{\alpha\beta} \quad (24)$$

The obvious factorization is

$$V^{\alpha}{}_{\mu}(x) = \delta^{\alpha}{}_{\mu} f(t) \quad (25)$$

with a consequent ambiguity about the sign. In the following (and hereafter, unless stated) it will be convenient to shift the origin of the normalized time to the boundary $t \to t - 1$, so

$$f^2 = a^2 = 1/t^2 \Rightarrow f = \pm 1/t \text{ or } f = \pm 1/|t| \quad (26)$$

In a spacetime with vierbein (25), and disregarding the relatively ineffective potentials near the boundary, the Dirac equation is therefore [12]

---

[10] Perhaps, if the first possibility is likened to a plane mirror in 3D, this possibility can be likened to a mirror surface matched to the phase of the radiation, which in any case conveys that in 1+1 D there is no distinction between the two possibilities.

[11] In all cases, though it may have appeared earlier that entropy was increasing, it will be clear at or near the boundary that the whole has remained perfectly synchronized all along.

$$i\hbar\gamma^\mu\partial_\mu\psi(x) - f(t)m\psi(x) = 0 \qquad (27)$$

We now sketch a method to resolve the ambiguity in (26), demanding that the wavefunction propagate across the boundary in the manner of the trajectories depicted in Fig. 4.[13] We will assume this means that

$$U\psi(-t,\mathbf{x};\pm e) = \psi(t,\mathbf{x};e) \qquad (28)$$

for some fixed linear or anti-linear operator $U$ that conserves the probability (including therefore the possibility of charge conjugation) and which is a generally respected symmetry of the dynamics.[14] Possibly relevant symmetries are charge and time - parity is excluded by (28). The form of Eq. (28) ensures the wavefunction spatial structure is mirrored about $t = 0$. Continuity *at* the boundary requires

$$U\psi(0,\mathbf{x};\pm e) = \psi(0,\mathbf{x};e) \qquad (29)$$

where for now we ignore the complication due to charge inversion. In the case in (26) that $f(t) = f(-t)$ the total operator in (27) will have the same time-parity as the Dirac equation in flat space, so both the time reversal and charge conjugation symmetry operations on the Minkowski-spacetime Dirac equation remain symmetries of (27). That is, if $\psi(t,\mathbf{x};e)$ is a solution of (26) then so are the following [6]

$$\begin{aligned}
\mathcal{T}\psi(t,\mathbf{x};e)\mathcal{T}^{-1} &= -i\eta_T\gamma^5 C\psi(-t,\mathbf{x};e) \\
\mathcal{C}\psi(t,\mathbf{x};e)\mathcal{C}^{-1} &= \eta_C C\gamma^{0*}\psi^*(t,\mathbf{x};-e) \\
\mathcal{TC}\psi(t,\mathbf{x};e)\mathcal{C}^{-1}\mathcal{T}^{-1} &= -i\eta_T\eta_C\gamma^5 C^2\gamma^{0*}\psi^*(-t,\mathbf{x};-e)
\end{aligned} \qquad (30)$$

where $C$ is a 4x4 matrix satisfying $C\gamma_\mu^T + \gamma_\mu C = 0$ and the $\eta$ are arbitrary phase factors. Let us now impose the condition that the two solutions either side of the boundary are equal. In the first case we get

$$\mathcal{T}\psi(t,\mathbf{x};e)\mathcal{T}^\dagger = \psi(t,\mathbf{x};e) \Rightarrow -i\eta_T\gamma^5 C\psi(-t,\mathbf{x};e) = \psi(t,\mathbf{x};e) \qquad (31)$$

This is a solution of (28) and (29) if

$$U = -i\eta_T\gamma^5 C, \quad (1 - i\eta_T\gamma^5 C)\psi(0,\mathbf{x};e) = 0\,\forall\mathbf{x} \qquad (32)$$

This has no solutions other than all four components of $\psi(0,\mathbf{x};e)$ vanish. The same applies to the third of (30). Therefore $f(t) = f(-t)$ is incompatible with the preservation of a respected symmetry across the boundary.

Observing that the alternative $f(t) = -f(-t)$ has the effect of changing the sign of the mass across the boundary, we introduce a mass-inversion operation $\mathcal{M}$ with the property that if $\psi(t,\mathbf{x};e,m)$ is a solution of Dirac's equation in Minkowski spacetime with mass $m$ then so is

$$\mathcal{M}\psi(t,\mathbf{x};e,m)\mathcal{M}^{-1} = \eta_M\gamma^5\psi(t,\mathbf{x};e,-m) \qquad (33)$$

with mass $-m$. Now let $f(t) = f(-t)$ and let the post-boundary particle have negative mass. Then $\psi(t,\mathbf{x};e,m)$ and

$$\begin{aligned}
\mathcal{MT}\psi(t,\mathbf{x};e,m)\mathcal{T}^{-1}\mathcal{M}^{-1} &= -i\eta_T\eta_M C\psi(-t,\mathbf{x};e,-m) \\
\mathcal{CMT}\psi(t,\mathbf{x};e,m)\mathcal{T}^{-1}\mathcal{M}^{-1}\mathcal{C}^{-1} &= -i\eta_T\eta_M\eta_C C\gamma^{0*}C^*\psi^*(-t,\mathbf{x};-e,-m)
\end{aligned} \qquad (34)$$

are all solutions of (27), whilst (29) must be generalized to

$$U\psi(0,\mathbf{x};\pm e,\pm m) = \psi(0,\mathbf{x},e,m) \qquad (35)$$

There are no non-trivial solutions of (35) for $U = C$. In the Dirac representation

$$C = C^*, \quad C^2 = -1, \quad \gamma^0 = \gamma^{0*} = \gamma^{0T} \Rightarrow C\gamma^{0*}C^* = C\gamma^0 C = -C^2\gamma^0 = \gamma^0 = \begin{pmatrix} I & 0 \\ 0 & -I \end{pmatrix} \qquad (36)$$

so (35) has non-trivial solutions if

$$U\psi \equiv \eta C\gamma^{0*}C^*\psi^* \qquad (37)$$

and either the upper pair of components (the 'large component') or lower pair of components (the 'small component') of the Dirac wavefunction vanishes. A more detailed analysis via solution of the Dirac equation (27) with $f = 1/t$ in terms of Hankel functions confirms that the small part decays towards the boundary, consistent with the effects of Hubble drag and the final zero speed fate of the classical trajectory.

In summary, the paths in Fig. 4 will be followed by quantum matter if the vierbein is

$$V^\alpha{}_\mu(x) = \delta^\alpha{}_\mu a(t) \tag{38}$$

and provided it can be arranged so that the wavefunction arriving at the boundary is such that only one of the two Dirac spinors is non-zero.[15] In that case the image in the time-like mirror will be a time-reversed, charge-conjugated, negative mass solution of the Dirac equation. Note that charge is conserved for as long as world lines are indexed by a monotonic time across the boundary; the charge conjugation applies to the world-line of the image, parsed in a reverse sense in time. If the pre-boundary particle is an electron for example, then the image (in forward time) is also an electron. This same image parsed in negative time is a positron with negative mass - consistent with the Stueckelberg interpretation (7).

---

[12] The mass is normalized to the Hubble constant, consistent with the normalization of the coordinates.
[13] We do not prove the assertions about the non-viability of the other candidates considered, nor show more generally that the found solution is unique.
[14] By 'generally respected' is meant a symmetry that applies well within lower half-space of the conformal system where, in case it becomes an issue, the effects of cosmological curvature are negligible.

## CPT Invariance

The foregoing is adequate for a QED universe because therein time reversal symmetry is respected. The Friedmann equations and the image universe remain synchronized and consistent either side of the boundary. But time reversal symmetry is not respected in the decays of neutral kaons. CPT is the only extension that is universally respected by the dynamics.(8) The association discussed here between the particle symmetry and global topology suggests it might be worth looking for a more faithful symmetry boundary at a singular point in the solution of Friedmann equation at which the post boundary 'image' is (also) a parity inverse of the pre boundary universe. This is to consider that the topology implied by the metric (3) and Fig. 4 is incorrect and perhaps other forms of the de Sitter spacetime with future singularities will suggest the correct topology. Two possibilities are briefly mentioned here.

Corresponding to the normalized but 'un-shifted' spacetime with line element

$$ds^2 = dx^2 / (1-t)^2 \tag{39}$$

the de Sitter spacetime can also be written [9]

$$ds^2 = dx^2 / (1 - x^2/4)^2 \tag{40}$$

Then a vierbein

$$V^\alpha{}_\mu(x) = \delta^\alpha{}_\mu / (1 - x^2/4) \tag{41}$$

gives a Dirac equation

$$\left(i\hbar\gamma^\mu \partial_\mu - \frac{m}{1 - x^2/4}\right)\psi(x) = 0 \tag{42}$$

The transformation relating the two and which is faithful at the origin is

$$t \to \frac{4t + 2(t^2 - r^2)}{(2+t)^2 - r^2}, \quad r \to \frac{4r}{(2+t)^2 - r^2} \tag{43}$$

which puts the conformal horizon studied earlier coincident with a *pair* of surfaces

$$t = 1 \to t = \pm\sqrt{r^2 + 4} \tag{44}$$

Consider then the pair of points

$$\{p^\mu\} = \left(\left|\sqrt{\mathbf{p}^2 + 4}\right| - |\varepsilon|, \mathbf{p}\right), \quad \{\bar{p}^\mu\} = \left(-\left|\sqrt{\mathbf{p}^2 + 4}\right| + |\varepsilon|, -\mathbf{p}\right) \tag{45}$$

approaching the positive-time surface from below, and is a 'complimentary' point that is similarly close to the negative-time surface, but approaching from above. Though the latter is parity inverted compared to the former, both points converge, for all **p**, to the same point on the boundary $t = 1$ in the earlier geometry as $\varepsilon \to 0$. This suggests that the upper half-space in the old system be identified with parity inverted negative times in the new chart. It remains to be shown though that this makes sense for the *propagation* of the wavefunction 'across' the boundary.

In a search for a CPT-compliant image the topology implied by the de Sitter metric expressed as the line element [10]

$$ds^2 = \sec^2 t \left(dt^2 - d\chi^2 - \sin^2\chi \, d\Omega^2\right) \tag{46}$$

---

[15] In practice this constraint on the spinor does not imply a reduced freedom to specify a wavefunction, locally, in the lower half-space, since the small component tends to zero automatically with the expansion.

may also be worthy of consideration. The analysis turns out to be complicated however by the fact that the metric is conformal to the Einstein static universe rather than Minkowski spacetime, and will not be considered further here.

## ELECTROMAGNETIC FIELDS

As for both classical and quantum matter there are two possible ways in which electromagnetic radiation may satisfy the GR-originated requirement for symmetry at the conformal boundary. For each of these and given a source at the origin, Figure 5 shows the response required from an image at $(t, \mathbf{x}) = (2, \mathbf{0})$. Just as in the case of matter, one of the alternatives (Fig. 5a) connotes the 'generation' at the boundary of a secondary, advanced, wave.

The majority of the present EM radiation energy density is in the CMB. Since the CMB is already thermalized and – more importantly here – statistically isotropic and presumed to be homogenous, it carries no signature of being advanced or retarded.[16] The distinction is therefore meaningless. Therefore the (only) effect of the boundary on the CMB is to require that it is (already) organized so as to cause self-cancelation at $t = 1$.[17]

At a time when the steady-state Cosmology seemed a viable possibility Hoyle (see the remarks in [11]) determined that the total energy density of a steady-state (and therefore de Sitter Cosmology) due to *starlight* is of the same order as the CMB.[18] In our universe therefore it may be that the integrated total arriving at the conformal boundary will be significant compared with the CMB. For the process in Fig. 5a to be tested (and potentially falsified) by observation would require the technical ability to see the reflected image of compact (and so divergent) light sources. Note that the image of a local source would not be red-shifted because the red and blue shifts would be cancelled by the round trip from the mirror.

Consider now the possibility depicted in Fig. 5b showing diverging radiation spontaneously re-converging in forwards time. Note that the source and its image have external legs, implying the possibility of interaction with other charges. In practice this means that the advanced component of the (pre-boundary particle) Green's function is retained, whereas the retarded part is subject to a condition analogous to (28). In order to discuss this possibility further let us recall that the EM action in curved spacetime is conventionally [5]

$$I = -\int d^4x \sqrt{-g}\left(\frac{1}{4}F_{ab}F^{ab} + A_a j^a\right) \tag{47}$$

where

$$F_{ab} := A_{b;a} - A_{a;b} = \partial_a A_b - \partial_b A_a \Rightarrow F^{ab} = g^{ac}g^{bd}\left(\partial_c A_d - \partial_d A_c\right) \tag{48}$$

Variation of the potentials gives [19]

$$A^{b;a}{}_a - A^{a;\ b}_{\ a} = j^b \tag{49}$$

In the particular case of conformal spacetime it is useful, using (48), to re-write the action (47) as

$$I = -\int d^4x\left(\frac{1}{4}F_{ab}F_{bd}\eta^{ac}\eta^{bd} + A_a\sqrt{-g}\, j^a\right) \tag{50}$$

Whatever its form, the covariant divergence of the current must vanish

$$j^a{}_{;a} = 0 \Rightarrow \partial_a\left(\sqrt{-g}\, j^a\right) = 0 \tag{51}$$

It is also true that the covariant divergence of the Minkowski current $\bar{j}$ must vanish:

$$\partial_a \bar{j}^a = 0 \tag{52}$$

Provided the Minkowski spacetime current is proportional to the curved spacetime current, it follows that (50) can be re-written in terms of the former:

---

[16] Retarded and advanced radiation may be distinguished *geometrically*, if somewhat approximately, by respective association with more-or-less spherically converging and diverging phase fronts in forwards time.
[17] At the least, one expects the conventional interpretation of the thermodynamic arrow of time would then be seriously challenged by any reasonable definition of entropy of the potentials.
[18] Hoyle speculated that the development of Cosmology would have been much different had he computed, in advance of the discovery of the CMB by Penzias and Wilson, the effective temperature of a thermalized distribution from his estimate of the steady-state energy density of radiation.
[19] This prescription is not a unique generalization of the minimally-coupled EM action in Minkowski spacetime to curved spacetime though it has the benefit of preserving gauge freedom  Other reasons for this choice that are given by Misner, Thorne and Wheeler seem less persuasive.

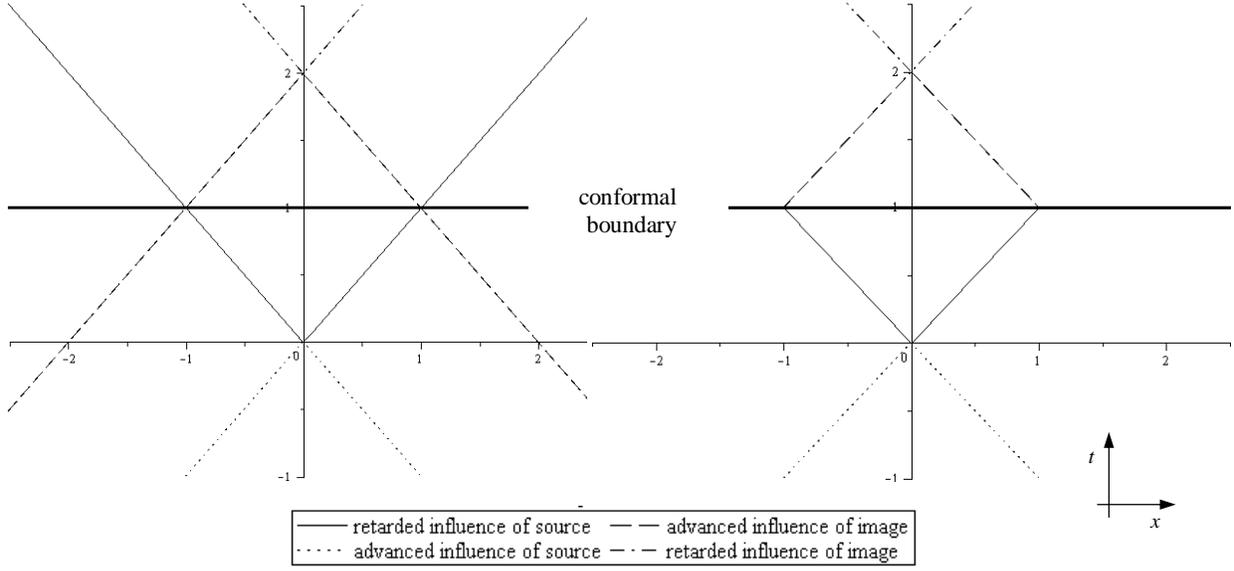

**Figure 5a.** The retarded light cone from a source at the origin is time-like reflected at the boundary generating 'advanced waves' in the pre-boundary half-space.

**Figure 5b.** The retarded light cone from a source at the origin is completely annihilated by its image.

$$I = -\int d^4x \left( \frac{1}{4} F_{ab} F_{bd} \eta^{ac} \eta^{bd} + A_a \bar{j}_b \eta^{ab} \right) \tag{53}$$

and now the scale factor is absent from the action. Variation of the covariant potentials in (53) must now give the Maxwell equations as if in Minkowski spacetime:

$$\partial^2 A_a - \partial_a (\partial \circ A) = \bar{j}_a \tag{54}$$

Independence from the scale factor is preserved if one chooses to work with the Minkowski spacetime Lorenz gauge,

$$\partial \circ A \equiv \eta_{ab} \partial_a A_b = 0 \tag{55}$$

leading to [20]

$$\partial^2 A_a = \bar{j}_a \tag{56}$$

Hence the curved spacetime covariant potential can be made the same as the Minkowski spacetime potential for a particular gauge choice.

One concludes from the above that the covariant potentials and Faraday tensor are *completely insensitive* to the scale factor, even as the scale factor passes through a singularity. Ordinarily, this would imply - due to gauge invariance - there can be no physical consequences of the boundary, and that radiation from a pre-boundary source must cross the boundary according to Fig 5a, with Fig. 5b ruled out. But here we will require continuity of the *potentials*, and boundary conditions on the potentials are not generally unchanged by a gauge transformation. This means that the boundary conditions can have a physical consequence that depends on the gauge.

Here, the boundary condition will be of the form (28), with the wavefunction replaced by some representation of the potential, in some gauge. The mirror symmetry of radiation must match that of matter, else the two would lose their synchronization, and the post boundary Cosmology could not evolve consistent with the Friedmann equation (7). Assuming the photon is massless we have [6]

$$\mathcal{CMT} A_\mu (t, \mathbf{x}; e) \mathcal{T}^{-1} \mathcal{M}^{-1} \mathcal{C}^{-1} = -A^\mu (-t, \mathbf{x}; -e) \tag{57}$$

For potentials to satisfy the mirror condition and remain solutions of the homogeneous Maxwell equations requires

$$A^\mu (-t, \mathbf{x}) = -\sigma A_\mu (t, \mathbf{x}) \tag{58}$$

---

[20] This is not a coordinate independent prescription to fix the gauge.

$\sigma$ is a sign freedom to accommodate the effect of $\eta$ in (37). Bearing in mind that the scale factor treats each side of (58) differently, one method of guaranteeing at least the implied mirror conversion of contravariant to covariant vectors – ignoring at first the time-reversed coordinate - is to employ a *covariant gauge* condition, e.g. as in the system [5]

$$A^{b;a}{}_a - A^{a;\ b}{}_a = 0, \quad A^{a;}{}_a = 0 \tag{59}$$

Now if $A_\mu(t, \mathbf{x})$ is a solution of this system then so is $A^\mu(t, \mathbf{x})$. (Note the same is not true of the condition (55) because the Minkowski-Lorenz gauge is not covariant - with the result that the covariant potentials either side of the boundary are not CMT mirror images of each other.) Then it remains only to ensure that the potentials are appropriately odd or even in time. $\sigma$ can be determined by considering the Coulomb field of a static charge in both half-spaces. The retarded field of the pre-boundary charge and the advanced field of the post-boundary image have the same sign because the symmetry operations are such that the charge retains its sign across the boundary *in forwards time*. Necessarily therefore

$$\phi(0_-, \mathbf{x}) = \phi(0_+, \mathbf{x}) \tag{60}$$

implying $\sigma = -1$ in (58). The general solution is therefore

$$\phi(-t, \mathbf{x}) = \phi(t, \mathbf{x}), \quad \mathbf{A}(-t, \mathbf{x}) = -\mathbf{A}(t, \mathbf{x}) \tag{61}$$

and in particular

$$\left.\frac{\partial \phi(t, \mathbf{x})}{\partial t}\right|_{t=0} = 0, \quad \mathbf{A}(0, \mathbf{x}) = \mathbf{0} \Rightarrow \mathbf{B}(0, \mathbf{x}) = \mathbf{0} \tag{62}$$

Eq. (62) should not be taken to imply that the boundary is a magnetic (super) conductor; there is no physical charge of any kind at the boundary capable of enforcing such a condition. Instead Eq. (62) implies a restriction on the dynamics of radiating charges elsewhere and away from the boundary; their motions must be such as to ensure the magnetic field vanishes on the boundary for example. These boundary conditions are to be applied to the Maxwell system in conformal spacetime with covariant gauge. The latter is

$$2\frac{\dot{a}}{a}\phi + \frac{\partial \phi}{\partial t} + \nabla \cdot \mathbf{A} = 0 \tag{63}$$

Eq. (63) is mandated here, removing gauge freedom from the traditional theory.

The Faraday tensor is insensitive to the gauge choice, so a classical theory need not respect (63). In fact in the classical domain one is free to ignore the gauge fixing (63) and return to the Minkowski space Maxwell theory (54) with full gauge freedom. One may choose for example the Lorenz gauge, and so compute fields from a potential satisfying (56). Even so, that 'neo-classical' theory will differ from the Cosmology-free Maxwell version through the (gauge-independent) requirement that the magnetic field vanish at the prescribed time.

Let us briefly consider the case of a static charge. Putting (63) into (54) gives the equations for the potentials:

$$\partial^2 \phi + 2\frac{\dot{a}}{a}\frac{\partial \phi}{\partial t} + 2\left(\frac{\ddot{a}}{a} - \frac{\dot{a}^2}{a^2}\right)\phi = \bar{\rho}, \quad \partial^2 \mathbf{A} = \bar{\mathbf{j}} + 2\frac{\dot{a}}{a}\nabla \phi \tag{64}$$

We will consider here only the de Sitter limit - written in Hubble units as $a = 1/t$. In that particular case the term in parentheses in (64) vanishes and (64) boils down to

$$\partial^2 \psi = e\delta^3(\mathbf{x})/t, \quad \partial^2 \mathbf{A} = -2\nabla \psi \tag{65}$$

where $\psi := a\phi$. A solution for $\psi$ with the correct parity and valid on both sides of the boundary is

$$\psi = \frac{\kappa}{2|\mathbf{x}|}\int dt' \frac{1}{t'}\left(\delta(t - t' - |\mathbf{x}|) + \delta(t - t' + |\mathbf{x}|)\right) = \frac{\kappa}{2|\mathbf{x}|}\left(\frac{1}{t - |\mathbf{x}|} + \frac{1}{t + |\mathbf{x}|}\right) = \frac{\kappa t}{|\mathbf{x}|x^2}; \quad \kappa \equiv \frac{e}{4\pi} \tag{66}$$

The potentials are then

$$\phi = \frac{\kappa}{|\mathbf{x}|}\frac{1}{1 - \mathbf{x}^2/t^2} \tag{67}$$

Using that

$$\nabla \psi = \kappa \hat{\mathbf{x}}(3 - t^2/\mathbf{x}^2)t/x^4 \tag{68}$$

a particular solution of (65) is found to be

$$\mathbf{A} = \kappa \hat{\mathbf{x}} t / x^2 \tag{69}$$

The corrections to the Minkowski form of the potentials are significant close to the boundary, and small in the present era ( t = - 1 in Hubble units). Inserted into the usual definition of the Faraday tensor, these expressions for the potentials give

$$\mathbf{E} = \kappa \hat{\mathbf{x}}/\mathbf{x}^2, \quad \mathbf{B} = \mathbf{0} \tag{70}$$

Though the potentials are 'non-Lorenzian', in the case of a static charge the boundary has no effect on the fields. The potentials are obviously different from the traditional Lorenz-gauge default however. The gauge-fixed expressions (67) and (69) suggest the possibility of direct detection, independently of the fields, through a quantum interference arrangement perhaps.

## RADIATION

The theory described above is perfectly time-symmetric about the boundary, but is not locally time-symmetric elsewhere. To demonstrate the latter it is instructive to consider the Green's functions for the sources. Exploiting the freedom in the *classical* theory to ignore (63) and (64), and deal instead with a Lorenz gauge, then at first one may write without prejudice

$$A_\mu = \frac{1}{2}(G_{adv} + G_{ret}) * j_\mu + f_\mu; \quad \partial^2 f_\mu = 0; \quad A_\mu(0,\mathbf{x}) = 0 \tag{71}$$

as a solution to (56), where $j_\mu$ is a source confined to the lower half-space. But we would prefer to have the particular integral and the complimentary function satisfy the boundary condition independently, so that the particular integral part can be considered a fixed feature of the sources in this theory. One solution (not unique) is obviously to write

$$f_\mu = \frac{1}{2}(G_{adv} - G_{ret}) * j_\mu + A_\mu^{(cf)} \tag{72}$$

in which case

$$A_\mu = G_{adv} * j_\mu + A_\mu^{(cf)}; \quad \partial^2 A_\mu^{(cf)} = 0, \quad A_\mu^{(cf)}(0,\mathbf{x}) = 0 \tag{73}$$

In the absence of a complimentary function, the potentials are exclusively advanced, as illustrated here in Fig. 6. The relative inversion compared to the Wheeler-Feynman outcome of exclusively *retarded* influences may be understood as due to the exchange of the Wheeler-Feynman boundary condition [12,13] - a perfect future absorber - for a perfect reflector (understood here in the sense of a phase-matched mirror).

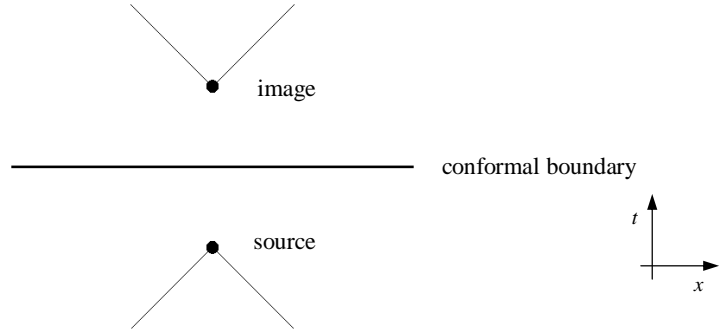

**Figure 6.** Cone of influence from a single event on the world line of a single source, with its image in the Conformal boundary.

As explained in the previous section, in the classical theory of the EM fields one may bypass (63) and (64), and work with the Lorenz gauge and (56). The effect of the boundary on the *homogeneous* fields is then expressed through the restricted Fourier expansion

$$\{A_\mu^{(cf)}\} = \int d^3k \, e^{-i\mathbf{k}.\mathbf{x}} \left( \hat{\mathbf{k}}.\mathbf{c}(\mathbf{k})\cos(|\mathbf{k}|t), i\mathbf{c}(\mathbf{k})\sin(|\mathbf{k}|t) \right) \tag{74}$$

Each mode is a standing wave in the extended space $t \in \mathbb{R}$. (Here $t = 0$ is the Cosmological boundary). Though the basis functions in time are restricted, they are still a complete set in just one (either one) of the two half-spaces. In a radiation gauge (74) would take the form

$$\phi = 0, \quad \mathbf{A} = \sum_{\lambda=1,2} \hat{\mathbf{e}}_\lambda \int d^3k \, c_\lambda(\mathbf{k}) e^{-i\mathbf{k}.\mathbf{x}} \sin(kt); \quad \hat{\mathbf{e}}_\lambda.\mathbf{k} = 0, \quad t \in \mathbb{R}_- \tag{75}$$

This is to be compared with the Maxwell default - with no boundary:

$$\phi = 0, \quad \mathbf{A} = \sum_{\lambda=1,2} \hat{\mathbf{e}}_\lambda \int d^3k \, e^{-i\mathbf{k}.\mathbf{x}} \left( b_\lambda(\mathbf{k}) \cos(kt) + c_\lambda(\mathbf{k}) \sin(kt) \right); \quad \hat{\mathbf{e}}_\lambda.\mathbf{k} = 0 \tag{76}$$

Generally, the effect of the difference between these two will be visible only at a distance from the boundary of order of the wavelength of interest.[21] Note however there is no cutoff in the ordinary sense.

The above pertains to classical field theory. Similar considerations apply also to QED and similar long wavelength effects predicted. To make QED comply with (73) it will probably be necessary to redefine the creation and annihilation operators and the ZPF, with special attention to the sign of the frequency.[22]

## TIME-ASYMMETRY

Wheeler and Feynman [12,13] are noted for their attempt to accommodate time-asymmetry within the direct action paradigm by appeal to 'boundary conditions'. The idea was to explain predominance of retarded over advanced radiation as due to the presence of future absorbers, extrinsically breaking the intrinsic time symmetry of the direct-action action. Implicitly their argument uses the second law of thermodynamics to explain the electrodynamic arrow of time, and has attracted dissent for putting thermodynamics above electrodynamics [14]. In any case the theory was ruled out by subsequent developments in Cosmology - the density of future absorbers required by the theory was found to be incompatible with admissible Cosmologies compatible with observation [15]. Though the Wheeler-Feynman *explanation* for broken symmetry was refuted by the absence of sufficient future absorbers, it will be useful here because had turned out that there *were* sufficient absorbers, there would have been no option but to accept the proposed explanation, since the mechanism and its predicted effect were not otherwise in doubt.

A reflecting boundary is a fixed property of the theory presented here. The outcome of a correspondingly modified Wheeler-Feynman argument – taking into account reflection rather than absorption – achieves here the status of certainty, granted veracity of this theory of course. But it is easy to show that the Wheeler-Feynman mechanism applied to a future reflector generates *anti-damping* radiation reaction, consistent with the advanced interaction associated with sources and derived in (73). It would seem to follow that the observational fact of retarded radiation and positive damping rule out the theory presented here. If so, that conclusion must apply to classical field theory, QED, and direct action, without distinction.

There is no conflict however if electromagnetic interactions on the advanced cone are principally *negative* rather than positive energy interactions. If indeed they were, then the emission of positive energy radiation on the retarded cone of a local source can be re-interpreted as an increment in the magnitude of negative binding energy propagating (in forwards time) along the (here, necessarily) advanced cone of that source. No future sinks or sources are then required. The predominance of retarded radiation as commonly understood then follows from the asymmetry of advanced Greens functions which are the consequence of the boundary condition associated with a future time-like mirror.

The strength required of the binding must be such as to accommodate the possibility that the all matter in question might be converted to radiation at any time, and therefore must at least equal the inertial mass. The only possibility is gravity; with some qualifications, on the large scale matter is gravitationally bound with an energy (magnitude) approximately equal to its mass. It appears that reconciliation of the symmetry of the Friedmann equation about the conformal singularity to electrodynamics is possible only if gravity is fundamentally electromagnetic.[23]

---

[21] The MOND threshold is at accelerations ~ $H$. Charged particles near light speed subjected to this acceleration generate wavelengths of this order, suggestive of a possible explanatory role for the boundary. In principle the boundary will affect the ZPF Fourier modes as in (75), perhaps with observable consequences manifesting through mass-renormalization.

[22] Possibly, it will be necessary to consider the Fourier basis states of the vacuum as negative energy oscillators associated with advanced radiation, the quantum states of which are 'occupied' by an infinite number of negative energy quanta. If so, then these quanta will be *destroyed* by the action of an operator that, in the traditional theory, would ordinarily create a photon.

[23] Suggestions along these lines restricted however to the direct action paradigm, have been made by the author elsewhere; recent numerical simulations have been promising.

# ACKNOWLEDGMENTS

I am very grateful to both Anthony Lasenby and Harold Puthoff for sharing their interest and expertise.